\documentclass[aps,nofootinbib,notitlepage,longbibliography,twocolumn, superscriptaddress]{revtex4-1}
\usepackage{amsmath,amssymb}
\usepackage{slashed}
\usepackage{graphicx}
\usepackage{bm}
\usepackage{float}
\usepackage[T1]{fontenc}
\usepackage{multirow}
\usepackage[utf8]{inputenc}
\usepackage{gauss} 
\usepackage[normalem]{ulem}
\usepackage[top=1.0in,bottom=1.0in,left=1.0in,right=1.0in]{geometry}
\usepackage[colorlinks,linkcolor=blue,citecolor=blue]{hyperref}
\usepackage{subfig}
\usepackage{booktabs}
\usepackage{caption}
\newcommand{\be}{\begin{equation}}
\newcommand{\ee}{\end{equation}}
\newcommand{\bea}{\begin{eqnarray}}
\newcommand{\eea}{\end{eqnarray}}
\newcommand{\ba}{\begin{eqnarray}}
\newcommand{\ea}{\end{eqnarray}}

\def\be{\begin{eqnarray}}
\def\ee{\end{eqnarray}}
\def\bea{\be}
\def\eea{\ee}

\def\roughly#1{\mathrel{\raise.3ex\hbox{$#1$\kern-.75em%
			\lower1ex\hbox{$\sim$}}}}

\usepackage{lipsum}

\date{\today}
\begin{abstract}
	We evaluate the entanglement entropy and entropic  function of  massive  two dimensional QED (Schwinger model) at finite temperature, density, and $\theta$-angle. In the strong coupling regime,  the  entropic function is dominated by the boson mass for large spatial intervals, and reduces to the CFT result for small spatial intervals.
	We also discuss the entanglement spectrum  at finite temperature and a finite $\theta$-angle. 
\end{abstract}
\begin{document}
	\title{Entanglement in massive Schwinger model\\
		at finite temperature and density}
	\author{Sebastian Grieninger}
	\email{sebastian.grieninger@stonybrook.edu}
	\affiliation{Center for Nuclear Theory, Department of Physics and Astronomy,
		Stony Brook University, Stony Brook, New York 11794–3800, USA}
	\author{Kazuki Ikeda}
	\email{kazuki.ikeda@stonybrook.edu}
	\affiliation{Center for Nuclear Theory, Department of Physics and Astronomy,
		Stony Brook University, Stony Brook, New York 11794–3800, USA}
	\affiliation{Co-design Center for Quantum Advantage (C2QA), Stony Brook University, USA}
	
	\author{Dmitri E. Kharzeev}
	\email{dmitri.kharzeev@stonybrook.edu}
	\affiliation{Center for Nuclear Theory, Department of Physics and Astronomy,
		Stony Brook University, Stony Brook, New York 11794–3800, USA}
	\affiliation{Co-design Center for Quantum Advantage (C2QA), Stony Brook University, USA}
	\affiliation{Department of Physics, Brookhaven National Laboratory
		Upton, New York 11973-5000, USA}
	
	\author{Ismail Zahed}
	\email{ismail.zahed@stonybrook.edu}
	\affiliation{Center for Nuclear Theory, Department of Physics and Astronomy,
		Stony Brook University, Stony Brook, New York 11794–3800, USA}
	\maketitle
	
	\section{Introduction}
	Quantum entanglement is the central tenet in the quantum description of all physical processes. In essence, quantum many-particle states are superposition states, which remarkably can lead to quantum correlations even when no explicit interaction is present. A quantitative way to capture this correlation is through the entanglement entropy.  
	
	Recently, there has been a renewal of interest in many aspects of quantum entanglement ranging from quantum many-body systems to field theory~\cite{Srednicki:1993im,Calabrese:2004eu,Calabrese:2009qy,Hastings:2007iok}, with new applications in quantum information science~\cite{Bauer:2022hpo}. 
	Of particular interest is the concept of quantum entanglement  flow, and its relation to quantum information flow, storage and encryption~\cite{Bekenstein:1981zz}.

	In nuclear and particle physics, quantum entanglement is inherent, yet the use of the entanglement entropy and its measurement has only started to receive attention recently. The quantum entanglement entropy encoded in the proton-proton (pp) interaction amplitude may be responsible for multiparticle production  observed at collider energies~\cite{Stoffers:2012mn}. Also, the quantum entanglement entropy  may be directly measurable in both pp
	and ep collisions at high energies ~\cite{Stoffers:2012mn,Kharzeev:2017qzs}.

	Quantum entanglement in two-dimensional non-conformal gauge theories with fermions has been investigated recently, with particular focus on  QED2~\cite{Ikeda:2023zil} and QCD2~\cite{THOOFT1974461}. 
	QED2 or the Schwinger model~\cite{Schwinger:1962tp} has been widely studied also as a testbed for quantum computation~\cite{Ikeda:2023vfk,PhysRevLett.131.021902,Ikeda:2020agk,PhysRevA.98.032331}.
	It has achieved remarkable notoriety given its solvability (in the massless case), and the non-perturbative aspects of its vacuum structure. Much like QCD4, the vacuum of QED2 exhibits a chiral anomaly, confines by generating a mass gap, exhibits a chiral condensate, theta-vacua and instantons. 
	
	In this work, we address the bosonized form of the Schwinger model in matter at a finite $\theta$-angle, in the strong coupling regime. We will use it to address the features of spatial entanglement, with a particular emphasis on the entropic function. We will first discuss the case of cold matter in detail, and then show how to introduce finite temperature.
	
	The organization of the paper is as follows: In section~\ref{SECA},
	we briefly review the bosonized form of QED2 in matter with a finite vacuum $\theta$-angle. The vacuum solutions for different fermion masses and vacuum angles exhibit modulated chiral waves. In section~\ref{SECB}, we derive the explicit form of the entropic function, which captures the UV-finite part of the spatial entanglement for a single spatial cut. The entropic function 
	is a universal function of the boson mass in QED2 shifted by the temperature and vacuum angle, with no effect from the finite density. It reduces to the known central charge in the CFT limit. In section~\ref{SECC}, we carry a numerical analysis of the
	entanglement entropy in QED2 at strong coupling, using the bosonized
	Hamiltonian. The eigenvalue spectrum of the entangled density matrix, is dominated by a collective  eigenmode that is sensitive to
	the boson mass. 
	Our conclusions are in section~\ref{SECE}. Some details are 
	given in the appendices.

	\section{Massive QED2 at finite density}
	\label{SECA}
	The Schwinger model with fermions of mass $m$ at finite chemical potential  $\mu$ is defined by~\cite{Schwinger:1962tp}
	\bea
	\label{A1}
	S=\int d^2x\,\bigg(\frac 14F^2_{\mu\nu} +\frac{\theta \tilde F}{2\pi}+\overline \psi (i\slashed{D}-m +\mu\gamma^0)\psi\bigg)
	\eea
	with $\slashed{D}=\slashed{\partial}-ig\slashed{A}$. The coupling $g$ has dimension of mass.
	The extensive interest in QED2 stems from the fact that it bears much in common with two-dimensional QCD (QCD2) and possesses confinement. As a result, the QED2 spectrum involves only chargeless composite excitations. Remarkably, the vacuum state is characterized by a non-trivial chiral condensate, and even topological tunneling configurations much like in QCD2. QED2, unlike QCD2, is exactly solvable in the massless case, a huge advantage in understanding its non-perturbative structure.

	Shifting the vacuum angle to the mass term and using the standard bosonization rules recalled in Appendix~\ref{app:bos}, the bosonized form of  (\ref{A1}) is readily obtained for small current masses:
	\begin{widetext}
		\bea
		\label{A2}
		S=\int d^2x \bigg(\frac 12(\partial_\mu\phi)^2 -\frac 12 f^2m_S^2\bigg(\frac\phi f-2{\mu x}\bigg)^2+f^2m_\pi^2
		\,{\mathbb N}_g{\rm cos}\bigg(\frac\phi f-2 \mu x-\theta\bigg)+\frac{\mu^2}{2\pi}\bigg)
		\eea
	\end{widetext}
	with the chirally shifted pseudo-scalar mass
	\bea
	m_S^2=\frac {g^2}\pi\qquad 
	m_\pi^2=-\frac{m\langle \overline{\psi}\psi\rangle_0}{f^2}
	\eea
	and  manifest periodicity in the vacuum angle. The x-dependent chemical potential in (\ref{A2}) follows from the bosonization rule 
	$$
	\frac 12 (\partial_\mu\phi)^2 -2\mu f\partial_x\phi=
	\frac 12 (\partial_\mu(\phi+2f\mu x))^2 +\frac{\mu^2}{2\pi}
	$$
	followed by the shift $\phi+2f\mu x\rightarrow \phi$.
	The origin of $m_S$ is anomalous (Schwinger ${\rm U}_A$(1) anomaly), and the chiral mass shift  $m_\pi$ originates from the emergent vacuum chiral condensate. Note that there is no Goldstone mode, owing to the Mermin-Wagner theorem.
	The shifting of the vacuum angle to the mass term follows from the Fujikawa construction,
	and dependence on $\theta$ disappears in the chiral limit. 
	The vacuum chiral condensate 
	$\langle\overline\psi \psi\rangle_0=-\frac{e^{\gamma}}{2\pi}m_S$
	is fixed using functional techniques (for example, using a torus as a regulator);
	$\gamma$ is the Euler constant~\cite{Sachs:1991en,Steele:1994gf}.   Like in QCD4, the chiral condensate is finite in the chiral limit. In Appendix \ref{LFX}, we discuss some subtleties of the chiral condensate on the light front.

\begin{figure}[H]
		\centering
		\includegraphics[width=\linewidth]{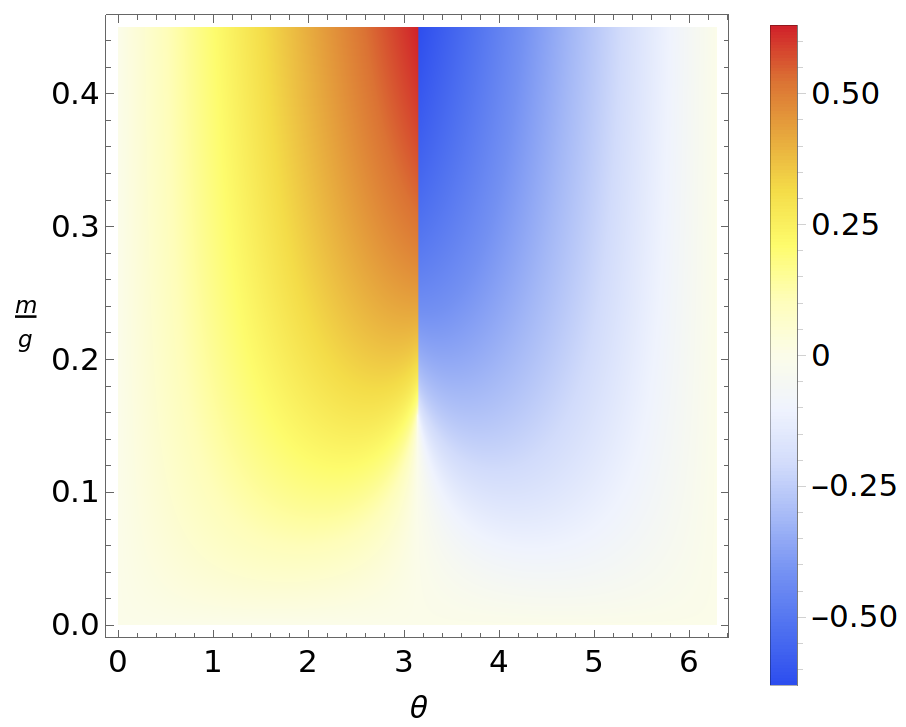}
		\caption{Heatmap of the general vacuum solution $\phi_v$ at zero temperature solution to \eqref{A4}. For $\theta=\pi$ and $\frac mg> 0.158$, the vacuum solution is degenerate, with a positive and negative solution. Only the positive solution is shown.}
		\label{fig:vacuum}
	\end{figure}

	The effective potential following from (\ref{A2}) is
	\bea
	\label{A3}
	&&\mathbb V(\phi, \theta, \mu) =\\
	&&\frac 12 m_S^2\bigg(\frac \phi f-2\mu x\bigg)^2-m_\pi^2
	\,\mathbb N_g {\rm cos}\bigg(\frac\phi f-2\mu x-\theta\bigg).\nonumber
	\eea
	The vacua are fixed by the minima solutions to the transcendental equation
	\bea
	\label{A4}
	{\rm sin}\bigg(\frac{\phi_v}f-2\mu x-\theta\bigg)
	=-\frac 1{\alpha} \bigg(\frac {\phi_v}f-2\mu x\bigg)
	\eea 
	with the dimensionless chiral parameter $\alpha=m_\pi^2/m_S^2$.
	The vacua solutions to (\ref{A4}) are shown in Fig.~\ref{fig:vacuum}. 
	For $\theta=\pi$, the vacuum is doubly degenerate for $\frac mg> \frac {m*}g=1/(2 \sqrt{\pi}
	e^\gamma) \approx 0.158$,\footnote{Our result differs by a factor of 1/2 from the result in \cite{Byrnes:2002nv,Ohata:2023gru} which estimates the mass of a single boson excitation
		semiclassically by fitting a harmonic oscillator to the effective potential at $\phi=0$ and then linearly extrapolates to a vanishing boson mass.} where  
	$$m^*=\frac {m_S^2}{|\langle\overline\psi\psi\rangle_0|}.$$
	This critical value deduced from a mean field analysis in the strong coupling regime is smaller than the numerical value of approximately $0.33$ reported in~\cite{Hamer:1982mx}.

	For the case $\theta=0$ and $\mu \neq 0$, the vacuum solution is inhomogeneous in space, with $\phi_v/f=2\mu x$. The fermion density
	per length is
	\bea
	\label{A6}
	\label{DENSITYF}
	n_F=\langle\psi^\dagger \psi(x)\rangle_\mu =\frac 1{\sqrt\pi}\partial^1\phi_v=\frac{\mu}\pi\equiv 2\int_0^{k_F}\frac {dk}{2\pi}
	\eea
	with the Fermi momentum $k_F=\mu$ and pressure $P=\frac{k_F^2}{2\pi}$ as expected. Using the bosonization rules, 
	this inhomogeneous solution gives rise to standing chiral waves in two-dimensions, 
	\begin{widetext}
		\bea
		\label{A7}
		\langle \overline\psi \psi(x)\rangle (\mu, 0) &=&\langle \overline\psi \psi\rangle_0\,{\rm cos}(\phi_v/f)=
		\langle \overline\psi \psi\rangle_0\,{\rm cos}(2\mu x),\nonumber\\
		\langle \overline\psi i\gamma^5\psi(x)\rangle (\mu, 0) &=&\langle \overline\psi \psi\rangle_0\,{\rm sin}(\phi_v/f)=
		\langle \overline\psi \psi\rangle_0\,{\rm sin}(2\mu x).
		\eea
	\end{widetext}
	The particle-hole (Overhauser) pairing  dominates over the particle-particle (BCS) pairing  since the Fermi surface is reduced to two disjoint points located $2k_F$ apart.  This is captured by the oscillations in (\ref{A7}) with wave-number $2k_F=2\mu$. The same observations were made for QCD2~\cite{Schon:2000he}, and QCD4 at strong coupling~\cite{Park:1999bz,Rapp:2000zd}.
	
	For the case $\theta\neq 0, \pi$, 
	the general vacuum solution follows by shifting away the $x$-dependence $\phi_v/f\equiv 2\mu x +\varphi_v/f$
	in (\ref{A4}), with $\varphi_v$ solution to
	\bea
	\label{A4XX}
	{\rm sin}\bigg(\frac{\varphi_v}f-\theta\bigg)
	=-\frac 1{\alpha}\frac{\varphi_v}f
	\eea 
	For small $\alpha$ near the massless limit, 
	\bea
	\label{SMALL}
	\frac{\varphi_v}f=\alpha\,{\rm sin}\theta+{\cal O}(\alpha^2)
	\eea
	with no effect on the fermionic density  \eqref{DENSITYF}. However, the underlying chiral wave (\ref{A7}) threading the Fermi surface from the Overhauser pairing, is modified
	\bea
	\label{A7X}
	\langle \overline\psi \psi(x)\rangle (\mu, \theta) &\sim &
	\langle \overline\psi \psi\rangle_0\,{\rm cos}\big(2\mu x + \alpha\, {\rm sin}\theta\big)\nonumber\\
	\langle \overline\psi i\gamma^5\psi(x)\rangle (\mu, \theta) &\sim &
	\langle \overline\psi \psi\rangle_0\,{\rm sin}\big(2\mu x+\alpha\, {\rm sin}\theta\big)\nonumber\\
	\eea
	
	In general, there are multiple vacua. For $\theta=\pi$ the vacuum is doubly degenerate. 
	The general vacua solutions to (\ref{A4XX}) for finite $\theta$ are shown in  Fig.~\ref{fig:vacuum}. The sign of the vacuum solution changes around the phase transition line at $\theta=\pi$. In the weak coupling phase with $\frac mg\gg 1$, the QED2 vacuum breaks C-symmetry or  $\varphi_v\rightarrow -\varphi_v$ symmetry. In the strong coupling phase with $\frac mg\ll 1$, the QED2 vacuum is C-symmetric with $\varphi_v=0$.

	\section{Entropic function}
	\label{SECB}
	We now consider space to be of length $L$ with a single cut of $l<L$, and study the spatial entanglement of $l$ with $L-l$. The space $L$ can be either periodic (close chain) or open (open chain). 
	For a single cut of length $l$, the spatial entanglement entropy can be analyzed using the replica construction~\cite{Calabrese:2004eu}. The UV
	insensitive entropic function is given by
	\bea
	C(l)=\frac{d{S_{EE}}}{d{\rm ln}l}.\label{eq:entrfu}
	\eea

	\begin{figure*}
		\centering
		\includegraphics[width=0.48\linewidth]{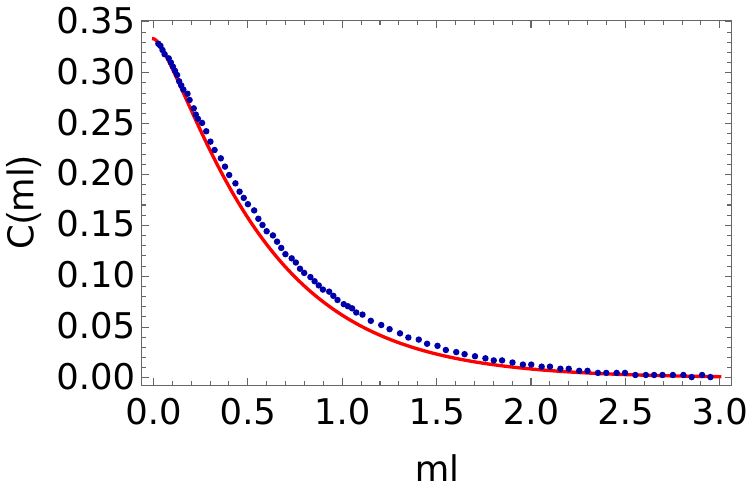}
		\includegraphics[width=0.48\linewidth]{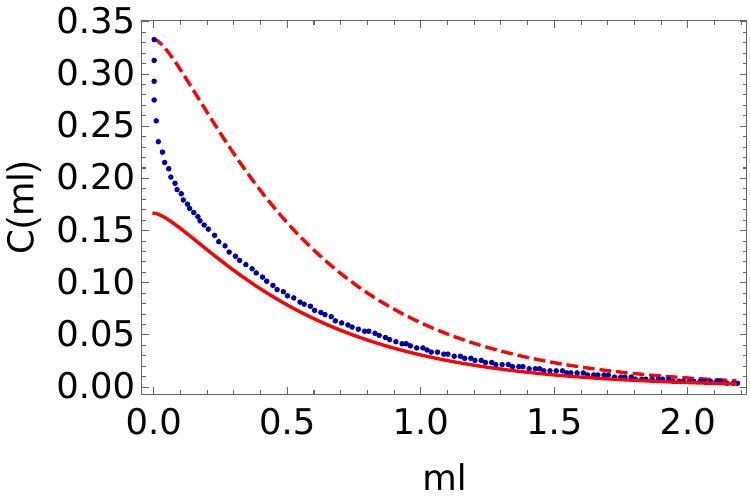}
		\caption{Left: entropic function for a massive fermion for periodic boundary conditions.  Right: entropic function for a massive boson for periodic boundary conditions. See text.}
		\label{fig:FF}
	\end{figure*}

	\subsection{Weak coupling regime: $\frac mg\gg 1$}

	In the weak coupling regime, the pseudo-scalar mass  is tachyonic,  
	$m_\theta^2\sim m_S^2-m_\pi^2<0$.
	The normal ordering with respect to the 
	pseudo-scalar mass $m_S\ll m$ is no longer justified for heavy fermions.
	In this regime, we revert to the case of free massive fermions in a screened phase, and use the newly developed field theoretical approach
	to the entanglement in~\cite{Liu:2022qqf} using replicated fermions. 
	
	In the absence of gauge interactions or screening, the entropic function is given in closed form by that of free massive fermions~\cite{Liu:2022qqf}
	\bea
	\label{CML1}
	C(ml)=\frac {ml}3\int_0^1 dx 
	\frac{\,K_1\bigg(\frac{ml}{\sqrt{x(1-x)}}\bigg)}{\sqrt{x(1-x)}}
	\eea
	which is seen to reduce to $C(ml\ll 1)\sim \frac 13$
	in the conformal limit, and asymptote
	\bea
	\label{FERASY}
	C(ml\gg 1)&\sim& \sqrt{\frac{\pi ml}6}\int_0^1 dx\,
	\frac{e^{-\frac{ml}{\sqrt{x(1-x)}}}}{(x(1-x))^{\frac 14}}\nonumber\\
	&\sim& \sqrt{\frac{\pi ml}3}\,e^{-2ml}.
	\eea
	(\ref{FERASY}) is consistent with the asymptotic form derived in~\cite{Casini:2005zv} using a Painleve V equation for the Renyi entropy, following from the  bosonization of free massive fermions and also \cite{Goykhman:2015sga}.
	In Fig.~\ref{fig:FF} (left)  we compare the central charge for the closed chain (\ref{CML1})  (solid-red curve) to  the exact numerical solution to the Painleve V analysis of the free massive fermion in~\cite{Casini:2005zv} (dotted-blue curve). The agreement  for the fermionic case is remarkable.
	
	In the presence of gauge interactions, the Renyi entropy and the ensuing entropic functions can be organized using standard Feynman graphs with $n$ replicated massive fermions. In the planar approximation, the resummed rainbow diagrams amounts to  a shift in the massive free fermion mass for QCD2~\cite{Liu:2022qqf}. 
	
	A rerun of these arguments for QED2 shows that 
	the dressing of the free massive fermion lines with rainbow diagrams yields
	also a shift
	\bea
	m\rightarrow \tilde{m}=\sqrt{m^2-m_S^2}
	\eea
	although the planar approximation is only valid for large replicas in QED2.
	The ensuing entropic function is still given by (\ref{CML1}), but with $m\rightarrow\tilde m$. 
	In particular, it reduces
	to the CFT value of $\frac 13$ for small
	intervals. For larger intervals, it asymptotes $e^{-2\tilde m l}$ with a range  
	$$l\approx \frac 1{2\tilde m} < l_S\approx \frac 1{m_S}$$  smaller than the screening range
	at weak coupling,
	thereby resolving the fermionic structure!

	For $m<m_S$ the shifted mass is tachyonic, a signal  that the weakly
	interacting phase undergoes a transition to a strongly interacting phase with bound fermions, with a critical value  $\frac {m_*}g=\frac 1{\sqrt{\pi}}>\frac 1\pi$ in the rainbow approximation.

	In general, the effects of temperature in QED2 on a heavy fermion have been analyzed using the invariant fermion propagator in~\cite{Steele:1994gf}. For time-like propagation, the bare fermion mass is shifted by the Coulomb self-energy $m\rightarrow m+\frac {\pi m_S}4$. The temperature correction drops out in the large time limit. For  space-like propagation, the fermion is screened with a screening mass $\frac{\pi T}2$. We conclude that for a heavy fermion, the result for the entropic function at finite temperature remains (\ref{CML1}) after the substitution  $m\rightarrow m+\frac{\pi T}2$ (space-like cut) and 
	$m\rightarrow m+\frac{\pi m_S}4$ (time-like cut).

	\subsection{Strong coupling regime: $\frac mg\ll 1$}

	\subsubsection{No matter: $\mu =0$, $T=0$}
	At strong coupling, the vacuum is C-even 
	and the screening is best captured in the bosonized form of QED2. 
	The case of a free massive bosons of mass $m_B$ was addressed using the Painleve V analysis, with  an  entropic function that asymptotes~\cite{Casini:2005rm}
	\bea
	\label{B2}
	C(m_Bl)\sim \frac{m_Bl}4\,K_1(2m_Bl),\qquad m_Bl\gg 1.
	\eea
	For small intervals, the entropic function 
	converges to the CFT value of $\frac 13$, in support of the boson-fermion
	duality in 2-dimensions.
	
	In Fig.~\ref{fig:FF} (right)  we compare  the central charge for a closed chain as given by  {\it half} of 
	(\ref{CML1}) (solid-red curve), to  the exact numerical solution to the Painleve V analysis for the free massive boson in~\cite{Casini:2005rm} (dotted-blue curve).  
	The full-Dirac  result  (\ref{CML1}) (dashed-red curve) is shown for comparison. 
	The overall agreement of the
	$\frac12$-Dirac fermion with the boson is also remarkable, except near the origin, where the massless bosonic fluctuations 
	become dramatically large for small intervals.

	A transition between the weak coupling regime with resolved fermions within the screening cloud, and the unresolved screened fermion as a boson in the strong coupling regime, can be differentiated  by  the entropic function.

	\subsubsection{Finite density}
	Using the decomposition $\phi=\phi_v+\xi$ around the vacuum solution (\ref{A4}) in the bosonized form (\ref{A2}),   we can
	infer the pertinent  bosonic mass for the entropic function. 
	More specifically, the small pseudo-scalar and axion fluctuations of (\ref{A3}) with fixed boundary variations $\delta_B\xi=\delta_B\vartheta=0$, are coupled both in vacuum and matter,
	\begin{widetext}
		\bea
		\big(\Box +m_S^2\big)\frac\xi f +m_\pi^2(T)
		{\rm cos}\bigg(\frac{\phi_v}f+\theta\bigg) \bigg(\frac \xi f+\vartheta\bigg)&=&0,\nonumber\\
		\Box \,\vartheta +m_\pi^2(T)
		{\rm cos}\bigg(\frac{\phi_v}f+\theta\bigg) \bigg(\frac \xi f+\vartheta\bigg)&=&0.
		\eea
	\end{widetext}
	For $\alpha\ll 1$, the pseudo-scalar field carries a squared mass
	\bea
	\label{MB3}
	m^2_\theta\sim m_S^2+m_\pi^2{\rm cos}\theta 
	\eea
	that is independent of $\mu$. 
	The pseudo-scalar meson disperses relativistically, despite the underlying inhomogeneous chiral wave! The Fermi surface consists only of two points $\pm k_F$ in phase space, with minor distortions in the vacuum pairing in the pseudo-scalar channel. This is also manifest from the sizeless light front wavefunction of the pseudo-scalar meson given in appendix~\ref{LFX}. With this in mind, 
	the asymptotic of the entropic function follows from (\ref{B2}) 
	with $m_B\rightarrow m_\theta$ or $C(m_\theta l)$.

	\begin{figure*}
		\centering
		\includegraphics[width=0.48\linewidth]{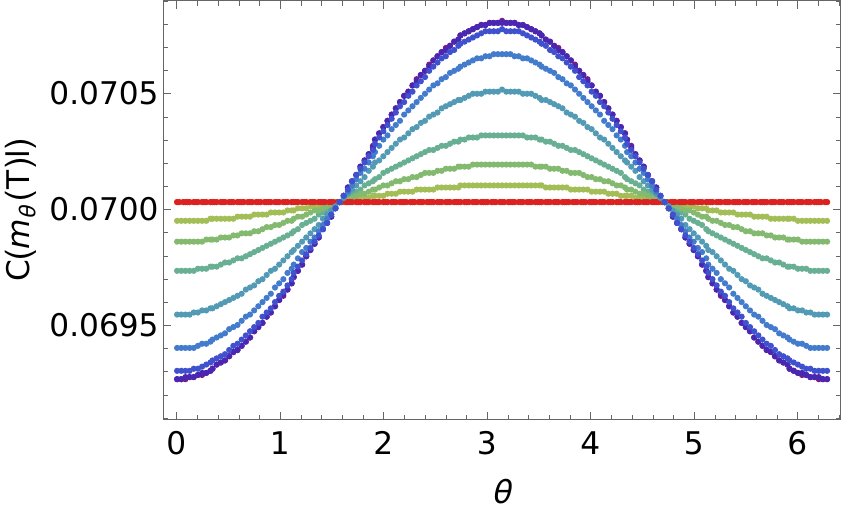}
		\includegraphics[width=0.48\linewidth]{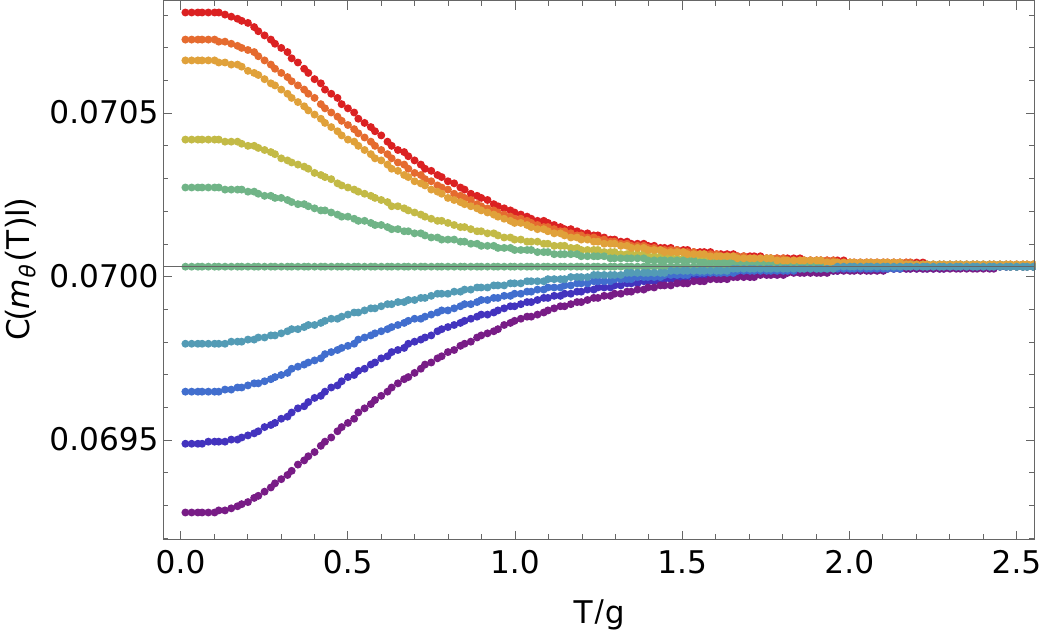}
		\caption{Entropic function in QED2 at strong coupling. Left:  as a function of the vacuum angle $\theta$, with increasing temperatures  from bottom-purple to top-red;  Right: as a function of temperature,  with increasing vacuum angle  $\theta=\frac \pi{20}$ bottom-purple, to $\theta=\pi$ top-red.
			We have set $m/g=0.0589$ with $l=1$ in units $g=1$. 
		}
		\label{fig:enttheta}
	\end{figure*}
	
	\begin{figure*}
		\centering
		\includegraphics[width=0.48\linewidth]{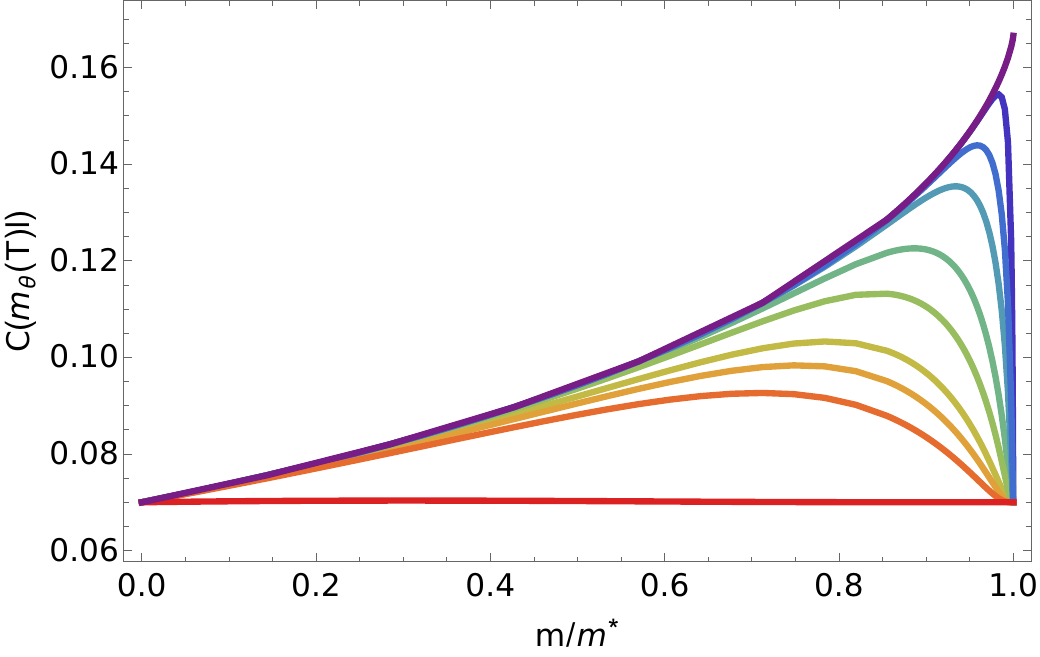}
		\includegraphics[width=0.48\linewidth]{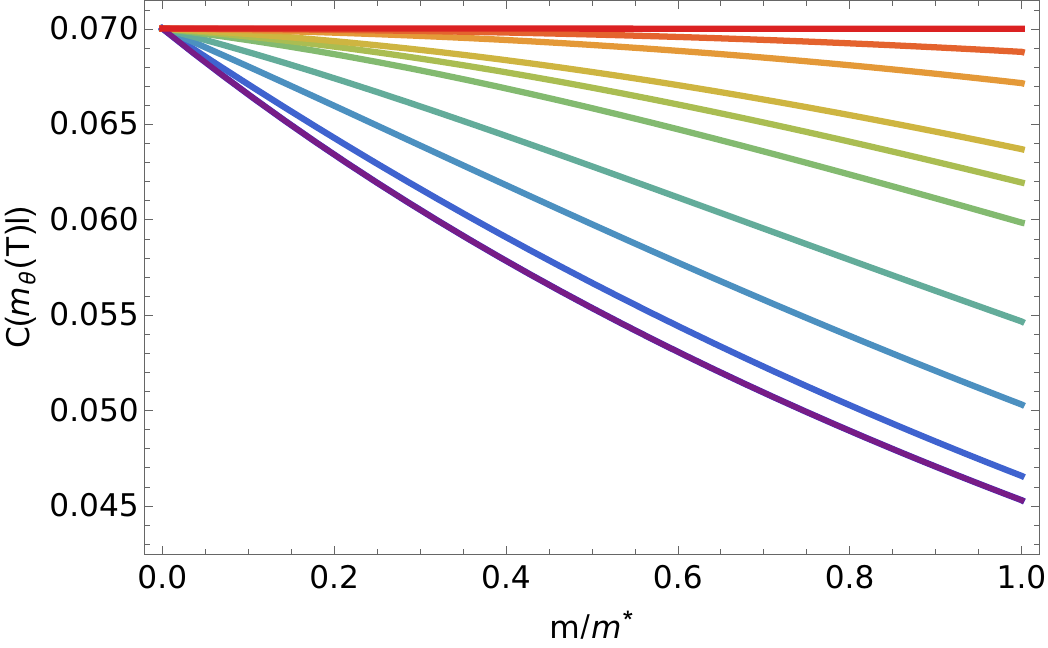}
		\caption{Entropic function versus the current quark mass $\frac m{m^*}$ for increasing temperatures in the strong coupling phase of QED2:  top-purple to
			bottom-red with $\theta=\pi$ (left), 
			and bottom-purple to top-red with $\theta=0$ (right). The interval length is $l=1$ in units with $g=1$.
		}
		\label{fig:entms}
	\end{figure*}

	\subsubsection{Finite temperature}
	The net effect of the temperature in the massive Schwinger model can be captured by an overall shift of the boson mass through the chiral condensate. 
	Near the massless limit, the boson mass is temperature-dependent, through the chiral condensate
	\bea
	m^2_\theta\rightarrow 
	m_\theta^2(T)\sim m_S^2+m_\pi^2(T)
	{\rm cos}\theta
	\eea
	with the thermal chiral condensate~\cite{Sachs:1991en,Smilga:1992hx,Fayyazuddin:1993ua,Steele:1994gf,Steele:1995sk,Durr:1996im}
	\bea
	\label{CONF}
	\langle\overline \psi\psi\rangle_T=
	\langle\overline \psi\psi\rangle_0
	e^{-2\pi\tilde\Delta_T(x=0)}
	\eea
	The vacuum-subtracted thermal massive boson propagator is
	\bea
	\tilde\Delta(x)=\int \frac{d^2k}{(2\pi)^2}
	e^{-ikx}\bigg(\frac {2\pi\delta(k^2-m_\theta^2)}{e^{\frac{|k_0|}T}-1}\bigg).
	\eea
	Note that (\ref{CONF}) can also be regarded as the resummed finite temperature tadpole contribution to the fermionic condensate, stemming from the subsumed normal ordering of the cosine in the bosonized
	form in (\ref{A2}).
	
	From (\ref{CONF}), we conclude that the entropic function  for QED2 in dense matter  both at finite temperature and density is
	\bea
	\label{MB2}
	\frac{lm_\theta (T) }4\,K_1(2lm_\theta (T))\qquad lm_\theta(T) \gg 1
	\eea
	with the low temperature  $T/m_\theta\ll 1$ behavior
	\bea
	m_\theta^2(T)\sim m_S^2+m_\pi^2
	{\rm cos}\theta
	\bigg(1-\sqrt{\frac {2\pi T}{m_\theta}}e^{-\frac{m_\theta}{T}}\bigg)\nonumber\\
	\eea
	and the high temperature  $T\gg g^2/\pi$ behavior
	\bea
	m_\theta^2(T)\sim m_S^2
	-\frac{m}{f^2}
	{\rm cos}\theta \,2Te^{-\frac{\pi T}{m_\theta}}.
	\eea
	Note that the chiral condensate has melted.

	\begin{figure*}
		\centering
		\includegraphics[width=0.48\linewidth]{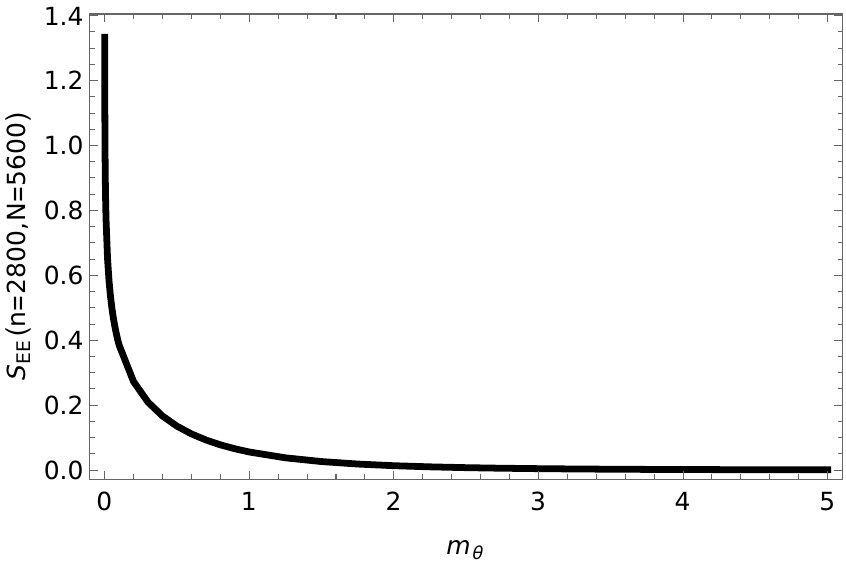}
		\caption{Entanglement entropy for the open chain versus the boson mass $m_\theta$ in units of $g=1$.}
		\label{fig:SEEMASS}
	\end{figure*}

	In Fig.~\ref{fig:enttheta} (left) we show the entropic function for a boson as $\frac 12$-Dirac fermion (\ref{CML1}), as a function of the vacuum angle $\theta$, for increasing temperatures from bottom-purple to top-red. The dependence on the vacuum angle disappears at high temperature, following the vanishingly small chiral condensate. 
	The interval size is set to $l=1$ in units where $g=1$, in the strong coupling regime $\frac mg=0.0589$. The high temperature limit $C(m_Sl)\sim 0.070$ corresponds to 
	a 'large interval' as per the right panel of Fig.~\ref{fig:FF}. 
	Fig.~\ref{fig:enttheta} (right) is an enlargement in the high temperature limit, 
	where the dependence on the vacuum angle disappears following the vanishing of the vacuum chiral condensate at high temperature. For  $\theta=\frac \pi 2, \frac {3\pi}2$, there is
	no dependence on the vacuum angle, as the 
	thermal chiral condensate drops out (the 'pion mass' contribution vanishes). 
	
	In the right panel of Fig.~\ref{fig:enttheta} we show the same entropic function versus temperature, with the vacuum angle increasing from $\theta=\frac \pi{20}$ (bottom-purple) to
	$\theta=\pi$ (top-red). For $\theta=\frac \pi 2$ the entropic function is independent of the temperature since the vacuum chiral condensate vanishes.
	
	In Fig.~\ref{fig:entms} we show again the entropic function versus the current quark mass $\frac m{m^*}$ with fixed vacuum angle $\theta=\pi$ (left) and $\theta=0$ (right), for increasing temperatures from $T=0$ (upper-purple) to 
	high temperature (bottom-red). 
	The critical mass $m^*$ is set by the tachyon condition or vanishing of the $\eta'$ mass at $\theta=\pi$
	$$m_S^2+m^*|\langle\overline\psi\psi\rangle|{\rm cos}\pi=0.$$
	At $\theta=0$, the entropic function dependence on the current mass ratio $\frac m{m^*}$ 
	drops out as the 'pion mass' vanishes exponentially.
	At $\theta=\pi$ and zero temperature, the entropic function reaches the CFT limit of $\frac 16$ at $m=m^*$. With increasing temperature, the CFT limit is never reached at $m=m^*$,
	which is a high temperature point
	whatever $T$ since the pseudoscalar mass $m_\theta$ ($\eta'$ mass) 
	vanishes. The entanglement entropy is maximal at the quantum critical point 
	with $\theta=\pi$ and $\frac {m_*}g$ as shown in Fig.~\ref{fig:SEEMASS}
	(CFT point with a massless boson), as noted initially in~\cite{Ikeda:2023zil}.

	\section{Entanglement density matrix at finite temperature}
	The entropic function indicates that in the regime of large $lm_\theta\gg 1$, massive QED2 in matter behaves like that of a free massive boson, while in the regime of small $lm_\theta\ll 1$, it resembles a CFT.
	To characterize the spectral properties of the  entangled density it is then sufficient to analyze the quadratic Hamiltonian from bosonized QED2.

	\begin{figure*}
		\centering
		\includegraphics[width=0.48\linewidth]{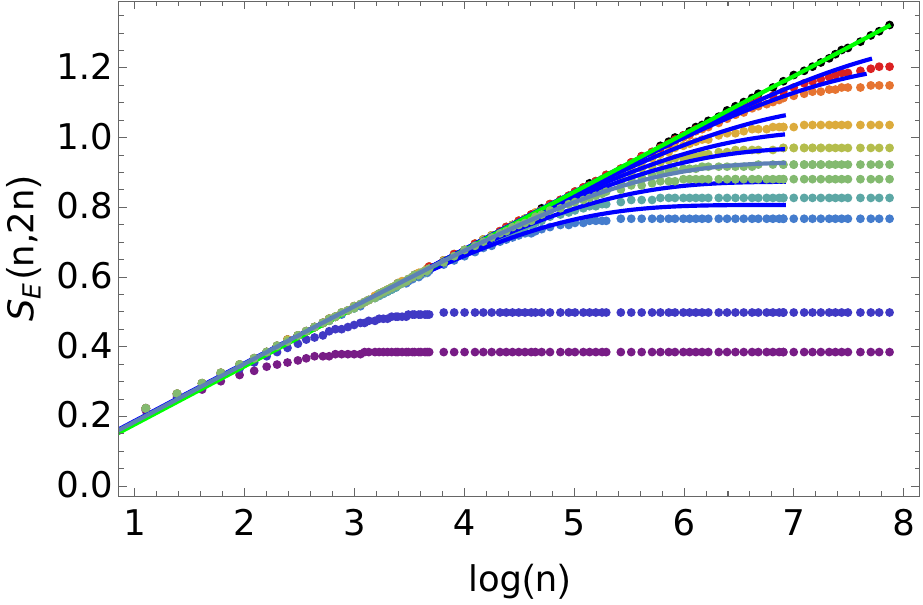}
		\includegraphics[width=0.48\linewidth]{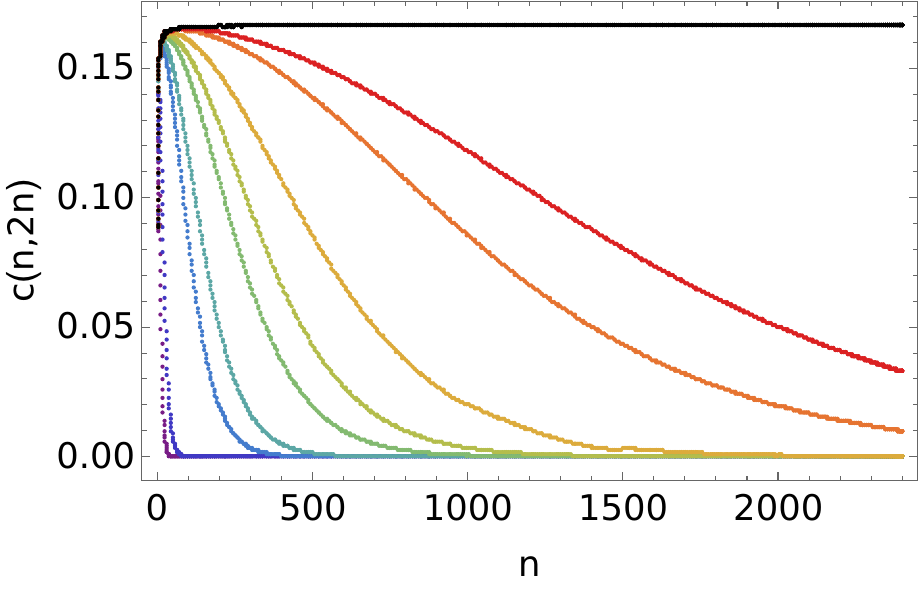}
		\caption{Left: entanglement entropy $S_E(n, N=2n)$ for the  open chain with different masses; Right: central charge for different masses, with the massless CFT limit $\frac 16$ shown in solid-black. See text.}
		\label{fig:SEE-MASS0}
	\end{figure*}

	\subsection{Chain Hamiltonian}
	\label{SECC}
	The small fluctuation Hamiltonian following from (\ref{A3}) is given by
	\bea
	\label{CONT}
	\mathbb H= \int dx 
	\frac 12 \big(\Pi^2 +{\xi^{\prime 2}}+{m_\theta^2(T)}\xi^2\big).
	\eea
	The boson mass is temperature and vacuum angle dependent from QED2. The  discretized form of (\ref{CONT}) is
	\bea
	\label{DIS1}
	\mathbb H\rightarrow\frac 12\sum_{i=1}^{N}
	\big(\Pi_i^2+(\xi_{i+1}-\xi_i)^2+a^2m_\theta^2(T)\xi_i^2\big)
	\eea
	with the chain spacing $a$ for large $l=na\,,L=Na$ but fixed 
	ratio $l/L$.
	(\ref{DIS1}) describes a discrete chain of coupled 
	oscillators with nearest neighbor couplings, with  
	either  open ($\xi_{N+1}=\xi_1=0$) or closed  
	($\xi_{N+1}=\xi_1$) boundary  conditions.
	(\ref{DIS1}) can be recast as
	\bea
	\label{DIS2}
	\frac 12 \sum_{i=1}^N \Pi_i^2+ 
	\frac 12 \sum^N_{i,j=1} \xi_i\mathbb K_{ij}\xi_j
	\eea
	with the real, symmetric and banded matrix
	\bea\mathbb K_{ij}=(2+a^2m_\theta^2(T))\delta_{ij}-\delta_{i,j+1}-\delta_{i, j-1}.
	\eea
	The ground state  of the chain (\ref{DIS1}-\ref{DIS2}) is 
	\bea
	\label{DIS3}
	\Psi[\xi]=\bigg(\frac{|{\Omega}|}{\pi^N}\bigg)^{\frac 14}
	e^{-\frac 12\sum_{i,j=1}^N\xi_i\Omega_{ij}\xi_j}
	\eea
	with energy $E=\frac 12 {\rm Tr}\,\Omega$.
	The covariance matrix $\Omega$ can be regarded as the square root of $\mathbb K$  under orthogonal rotations. The details of
	the diagonalization of $\Omega$ are given in Appendix~\ref{DETAILS}.

	\begin{table}[h]
		\begin{center}
			\begin{tabular}{l c c}
				\toprule
				\multirow{2}{3cm}{$m_\theta$  (eq. \eqref{CONT})		}		& \multicolumn{2}{c}{Parameters eq. \eqref{MAJORANA}}	\\
				\cmidrule{2-3}					&$m^c_\theta$	& shift $\Lambda'$	\\ \addlinespace
				\midrule
				$7\cdot 10^{-4}$		&0.00020&0.013\\ \addlinespace[2pt]
				$10^{-3}$		&	0.00031	&0.014\\ \addlinespace[2pt]
				$2\cdot10^{-3}$		&	0.00063&	0.015\\ \addlinespace[2pt]
				$3\cdot10^{-3}$		&	0.001& 0.0165\\\addlinespace[2pt]
				$4\cdot10^{-3}$		&	0.00139& 0.0174\\\addlinespace[2pt]
				$5\cdot10^{-3}$		&	0.00175& 0.0185\\\addlinespace[2pt]
				$7\cdot10^{-3}$		&	0.00245& 0.019\\\addlinespace[2pt]
				$10^{-2}$		&	0.0037& 0.021\\ \addlinespace[2pt]
				\addlinespace
				\bottomrule
			\end{tabular}
			\caption{Relation of mass \eqref{CONT} used in numerics to continuum formula \eqref{MAJORANA} in units of $a=1$.}\label{tab:1}
		\end{center}
	\end{table}

	\begin{figure*}
		\centering
		\includegraphics[width=0.49\linewidth]{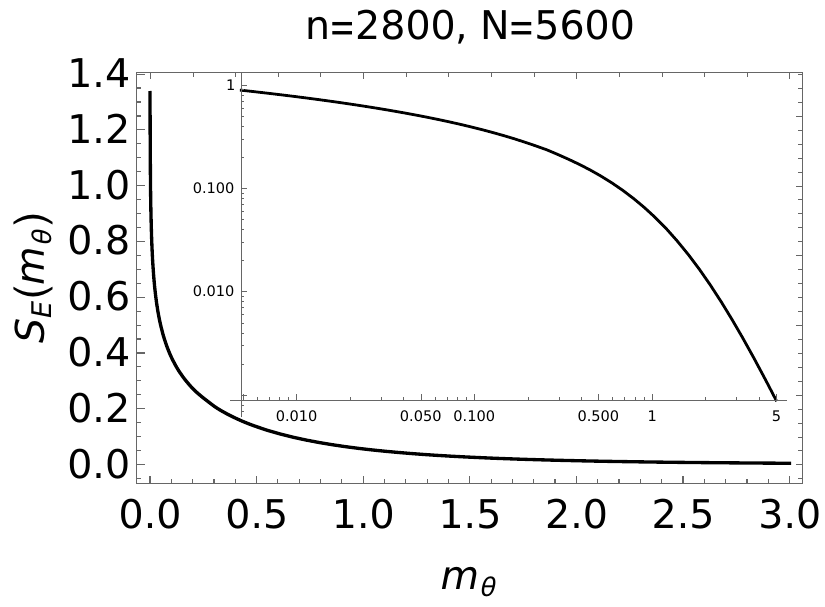}
		\includegraphics[width=0.49\linewidth]{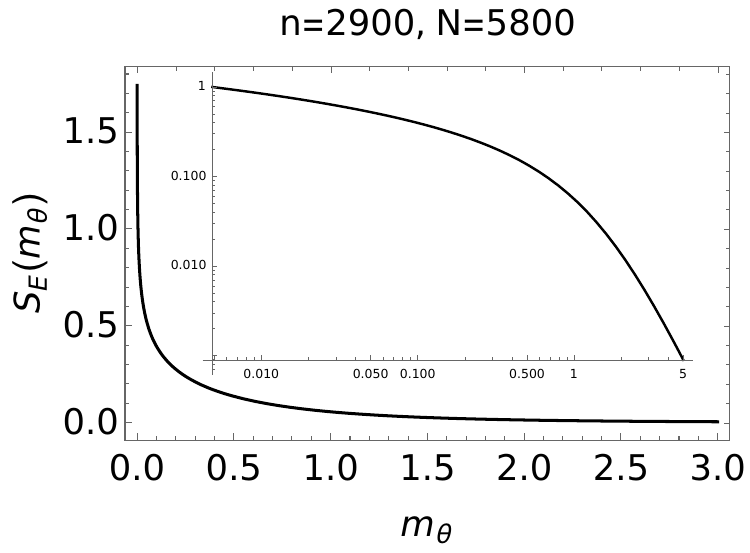}
		\caption{Left: EE for an open chain of size $n$
			versus $m_\theta$. Right: EE for a closed chain of size $n$ versus $m_\theta$. Both insets are double logarithmic plots.}
		\label{fig:SEE-nfixed}
	\end{figure*}

	\subsection{Entanglement  density matrix}
	The pure density matrix of the ground state of the open chain is given by
	\bea
	\label{DIS6}
	\rho[\xi, \xi']=\Psi[\xi]\Psi^T[\xi'].
	\eea
	The entanglement density matrix can be obtained by subdividing the chain $[\xi]\rightarrow [\underline\xi, \bar \xi]$
	\bea
	\label{DIS8X}
	\rho_E[\bar\xi, {\bar\xi}']=
	\sum_{\underline \xi} 
	\Psi[\underline \xi, {\bar \xi}]\Psi^T[\underline \xi, {\bar x}'].
	\eea
	The positive eigenvalues follow by diagonalizing  (\ref{DIS8X})
	\bea
	\label{DIS9X}
	\sum_{{\bar \xi}'} \rho_E[\bar\xi, {\bar\xi}'] \psi_l[{\bar \xi}']=p_l\psi_l[\bar\xi]
	\eea
	with the entanglement entropy 
	\bea
	S_{E}=-\sum_{l=0}^\infty p_l\,{\rm ln}\,p_l.
	\eea
	The implicit form of $p_l$ from the blocking of the covariance matrix $\Omega$ for the open and closed chains are given in Appendix~\ref{DETAILS}.

	\subsection{Entanglement Spectrum}
	We consider the Gram–Schmidt orthonormalization of the ground state
	\begin{equation}
	\Psi[\xi]=\sum_{i=1}c_i\Psi^A_i[\xi]\Psi^B_i[\xi],
	\end{equation}
	where $A$ and $B$ denote the non-empty disjoint subspaces such that $A\cup B=\{1,\cdots,N\}$. The entanglement spectrum is defined by the set of the coefficients $\{c_i\}$. In our previous paper~\cite{Ikeda:2023zil}, we showed numerically that the gaps in the entanglement spectrum close around the critical point.

	\subsection{Numerical results}
	For the blocked chain Hamiltonian, the explicit eigenvalues are given by (\ref{DIS11}), hence
	\begin{widetext}
		\bea
		\label{SEEX}
		S_{E}(n,N)=-\sum_{i=1}^n\sum_{l=0}^\infty 
		p_l[N,n,i]\,{\rm ln}\,p_l[N,n,i]
		=-\sum_{i=1}^n
		\bigg({\rm ln}(1-\lambda_{N,n,i})+
		\frac{\lambda_{N,n,i}}{1-\lambda_{N,n,i}}\,{\rm ln}\,\lambda_{N,n,i}\bigg)
		\eea
	\end{widetext}

	In Fig.~\ref{fig:SEE-MASS0}-left,
	we show the entanglement entropy results for the massless and open chain in $m_\theta=0$ (black). The green dashed line is \eqref{MAJORANA} with $m_\theta\to 0$ shifted by $+0.01(1)$. The curve can also be fitted with
	$$0.0108 + 0.1666\log(n)$$
	with the slope converging to the continuum limit of $\frac 16$.
	
	The rainbow colored lines are for a massive open chain with fixed $a=1$ and increasing masses  $m_\theta=\{7\cdot 10^{-4},10^{-3}, 2\cdot10^{-3}, 3\cdot10^{-3}, 4\cdot10^{-3},5\cdot10^{-3},7\cdot10^{-3},10^{-2},5\cdot10^{-2}, 10^{-1}\}$ (rainbow from red-top to purple-bottom). 
	They may be fitted by the continuum result \eqref{MAJORANA} by choosing the parameters indicated in table \ref{tab:1}. The masses are approximately related by $m^c_\theta\sim m_\theta^{1.2}$ and the shift scales approximately like $0.0044 \log(12.20 + 9810 \,m_\theta)$.

	In Fig.~\ref{fig:SEE-MASS0}-right, we show
	the numerical results for the corresponding central charge versus the interval length $na\equiv n$ for the open chain at different temperatures,  following from
	Fig.~\ref{fig:SEE-MASS0} (left) according to Eq.~\eqref{eq:entrfu} (the color coding of the curves corresponding to varying the mass is identical to Fig.~\ref{fig:SEE-MASS0}-left).
	For small intervals, the central charge is $\frac 16$ as expected from CFT, whatever the temperature. For large intervals, 
	the central charges is seen to increase from low temperature (solid-purple) to high temperature (solid-red) as the $\eta'$ mass becomes lighter. The black-solid curve is the CFT limit.
	
	In Figs.~\ref{fig:SEE-nfixed}, we show the EE for an open chain (left) and a closed chain (right) with fixed interval size $n$, in QED2 at strong coupling, as a function of the mass $m_\theta$. 
	The inset is a double logarithmic plot. The large logarithmic enhancement at small $m_\theta$, reflects on the CFT  limit. In the large mass limit, the fall off is exponential.

	\section{Conclusions}
	\label{SECE}
	
	We  have used the bosonized form of QED2 with a finite vacuum angle, to understand quantitatively the nature and 
	aspects of quantum entanglement in space.
	At weak coupling the QED2 vacuum breaks C-symmetry and chiral symmetry, with C-symmetry restored at strong coupling.

	When matter is added, a spatially inhomogeneous chiral density wave threading the Fermi-surface by Overhauser (particle-hole) pairing forms. The chiral condensate is depleted by thermal effects. The massive boson of the interacting theory carries a thermal mass that is only sensitive to the chiral condensate. 
	
	To quantify the spatial entanglement in QED2, we have traced out a spatial interval.  The spatial entanglement entropy  is a function of the length of this spatial interval, the temperature and the vacuum angle. For fixed temperature, the central charge increases 
	with vacuum angle. At high temperature, the central charge is insensitive to the vacuum angle, as the chiral condensate becomes exponentially suppressed.

	To check our mean field results numerically in the strong coupling regime,  we have made use of the bosonized Hamiltonian for QED2. 
	The discretized Hamiltonian is a  massive chain of coupled oscillators, with the mass encoding the combined effects of temperature and vacuum angle. 
	For open chains, the spatially entangled density matrix is characterized by an eigenspectrum, which is dominated by a large
	collective eigenvalue. The entanglement entropy asymptotes its CFT limit for small intervals whatever the mass. For large intervals, 
	the entanglement entropy flattens out with increasing mass. The
	corresponding central charge for short intervals is independent
	of the mass. For large intervals, it falls exponentially fast with the mass. This behavior is supported by our analytical analysis.
	
	QED2 is characterized by an axial anomaly and a chiral condensate, and could be regarded as a model for understanding the interplay of the chiral breaking and the axial-anomaly in QCD4
	in vacuum and at finite temperature. Our analysis of the central charge in QED2 at finite temperature and density shows that for large intervals, the central charge is very sensitive to the temperature and vacuum angle. At high temperature, 
	the central charge is totally controlled by the anomalous mass contribution to the $\eta'$ mass in the strong coupling regime. This observation may be used to disentangle
	the chiral and U$_A$(1) restoration in QCD4. 
	
	One natural extension of our results is to discuss the case of multiple flavors along the lines of~\cite{Banuls:2016hhv,Funcke:2022uwc,Dempsey:2023gib,Dempsey:2023fvm}; we will report on it in a forthcoming publication.

	\vskip 0.5cm
	{\noindent\bf Acknowledgments}
	
	\noindent 
	This work is supported by the Office of Science, U.S. Department of Energy under Contract No. DE-FG88ER40388 (SG, DK, IZ) and  DE-SC0012704 (DK), and the U.S. Department of Energy, Office of Science, National Quantum Information Science Research Centers, Co-design Center for Quantum Advantage (C2QA) under Contract No.DE-SC0012704 (KI, DK).
	
	\appendix
	\section{Bosonization relations}\label{app:bos}
	\noindent Our conventions for the bosonizations are
	those in~\cite{Coleman:1974bu}
	\bea
	\overline\psi\gamma^\mu\psi&=&\frac 1{\sqrt\pi}\epsilon^{\mu\nu}\partial_\nu\phi\nonumber\\
	\overline\psi\gamma^5\gamma^\mu\psi&=&\frac 1{\sqrt\pi}\partial^\mu\phi\nonumber\\
	\overline\psi\psi&=&\langle\overline\psi\psi\rangle_0\,
	\mathbb N_g{\rm cos}(\phi/f)\nonumber\\
	\overline\psi i\gamma^5\psi&=&\langle\overline\psi\psi\rangle_0\,
	\mathbb N_g{\rm sin}(\phi/f)
	\eea
	with $f=\frac 1{\sqrt{4\pi}}$ the analog of the meson decay constant, $\langle\overline\psi\psi\rangle_0$ the chiral condensate following from the normal ordering  $\mathbb N_g$ with respect  to the mass $m_S$.
	
	\section{Mass shift on the light-front}
	\label{LFX}
	\noindent The mass contribution in (\ref{MB3}) is expected, but with a naive  'mass dependent condensate' on the light-front.   Indeed, in the massless limit QED2 bosonizes to a massive boson of mass
	$m_S$. Its normalized partonic  light-front wavefunction $\psi_0(x)={\theta(x)}$
	defines the normalized Fock state
	\bea
	|B(p)\rangle =\int_0^1\frac{dx}{\sqrt{2x \bar x}}\psi_0(x) a^+(k)b^\dagger(p-k)|0\rangle
	\eea
	with $x$ referring to parton-x here.
	The mass term shifts the boson mass $\frac{g^2}\pi$ in first order perturbation theory by
	\bea
	\label{LF2}
	\frac{\langle B(p)|H_m|B(p)\rangle}
	{\langle B(p)|B(p)\rangle}=
	\int_0^1 dx \frac{m^2{\rm cos}\theta}{x\bar x}|\psi_0(x)|^2
	\eea
	Now we note that the vacuum fermion condensate on the light-front is
	\bea
	\label{LF3}
	\langle\overline\psi \psi\rangle_0=
	-m\int \frac{dk^+}{2\pi}\frac{\epsilon(k^+)}{k^+}
	=-\frac {m}{4\pi}\int_0^1\frac{dx}{x\bar x}
	\eea
	Inserting (\ref{LF3}) into (\ref{LF2}) yields
	\bea
	\label{LF4}
	\frac{\langle B(p)|H_m|B(p)\rangle}
	{\langle B(p)|B(p)\rangle}=
	-\frac{m\langle\overline\psi \psi\rangle_0}{f^2}\,{\rm cos}\theta
	\eea
	in agreement with (\ref{MB3}).
	Note that both (\ref{LF2}) and (\ref{LF3}) are IR sensitive and only defined modulo an IR regulator. However, the identity (\ref{LF4})  is independent of the regulator.
	This observation implies that the fermion condensate after proper IR regulator
	is m-independent, as per the IR regulated QED2 on the light front~\cite{Liu:2022qqf}, and the invariant torus regularization in~\cite{Sachs:1991en,Steele:1994gf}.

	\section{Block  diagonalization}
	\label{DETAILS}\noindent
	The covariance matrix  $\Omega=U^T \sqrt{K_D} U$ is the square root of  $\mathbb K=U^T K_D U$, which  is real symmetric, with $K_D$ referring to its diagonalized form. The eigenvalues and eigenvectors of $\mathbb K$  follow from Bloch's theorem
	\bea
	\label{DIS4}
	\lambda_k(\theta, T)&=&2+a^2m_\theta^2(T)-2{\rm cos}\bigg(\frac {\pi k}{2p+1}\bigg)\nonumber\\
	\alpha_k^l&=&\sqrt{\frac 2{2p+1}}\,{\rm sin}\bigg(\frac{\pi k l}{2p+1}\bigg)
	\eea
	The eigenvalues are labeled by $k$, and the eigenvectors
	by $k,l=1,2, ..., 2p=N$ for a chain with even links for simplicity. In particular for an open chain, the orthogonal matrix $U$ and the covariance matrix $\Omega$ are respectively  $U_{ml}=\alpha_m^l$ and
	\begin{widetext}
		\bea
		\label{DIS5}
		\Omega_{ml}=\bigg(\frac{2}{2p+1}\bigg)\sum_{k=1}^{2p} 
		{\rm sin}\bigg(\frac{\pi mk}{2p+1}\bigg)
		\bigg(2+a^2m_\theta^2(T)-2{\rm cos}\bigg(\frac{\pi k}{2p+1}\bigg)\bigg)^{\frac 12}
		{\rm sin}\bigg(\frac{\pi kl}{2p+1}\bigg)
		\eea
	\end{widetext}
	For a periodic or closed chain with $N=2p$
	\bea
	\alpha_k^l=\frac 1{\sqrt{N}}
	{e^{-i\frac{2\pi kl}N}}
	\eea
	and
	\begin{widetext}
		\bea
		\Omega_{ml}=\frac 1{N}\sum_{k=1}^N
		{e^{i\frac{2\pi mk}N}}
		\bigg(2+a^2m_\theta^2(T)-2{\rm cos}\bigg(\frac{2\pi k}{N}\bigg)\bigg)^{\frac 12}
		{e^{-i\frac{2\pi kl}N}}
		\eea
	\end{widetext}
	For $m_\theta=0$ the sum can be unwound, with the result at large $N$
	\bea 
	\Omega_{ml}\rightarrow \frac 1N\frac{{\rm sin}\frac \pi N}
	{{\rm cos}\frac {2\pi}N(m-l)-{\rm cos}\frac \pi N}
	\eea

	To construct the entangled density matrix and ensuing entanglement entropy we follow  the original construction in~\cite{Srednicki:1993im} with the recent application in~\cite{Liu:2018gae}.
	For an open chain, we fix the end points to $\xi_{N+1}=\xi_1=0$.
	Without loss of generality, we set $N=2p$ and subdivide the chain
	by using the labels
	\bea
	\label{DIS7}
	[N]=[n]\cup [N-n]
	\eea
	The entanglement between the split chain of size $[n]$ and the one with size $[N-n]$, can be calculated by blocking the covariance matrix $\Omega$ in (\ref{DIS5})
	\bea
	\Omega=
	\begin{pmatrix}
		A & B\\
		B^T& C
	\end{pmatrix}
	\eea
	with the rectangular matrix
	\bea 
	\label{DIS8}
	B_{N,n}=\Omega_{m\bar m}, \qquad m\in (1,...n),\qquad \bar m\notin (1,...n)\nonumber\\
	\eea

	\subsection{Case $n=1$}\noindent
	We now consider the simplest case of a periodic chain with $n=1$. The ground state wavefunction is
	\bea
	e^{-\frac 12 x^T\Omega_{N-1}x-x_1\beta^T x-x_1^2\frac 1{2N}{\rm cotan}\frac \pi{2N}}
	\eea
	with the vector entries $x_{a=1,.., N-1}$ and $\beta_{m=2, .., N}$
	\bea
	\beta_m= \frac 1N\frac{{\rm sin}\frac \pi N}
	{{\rm cos}\frac {2\pi}N(m-1)-{\rm cos}\frac \pi N}
	\eea
	The bi-variate density matrix is then
	\bea
	&&\rho[x_a,x_b]=\nonumber \\ 
	&&e^{-\frac{1}{2}x_a^{T}\Omega_{N-1}x_a-\frac{1}{2}x_b^{T}\Omega_{N-1}x_b-\frac{N}{4}\tan \frac{\pi}{2N}(\beta^{T}x_a+\beta^{T}x_b)^2}\nonumber\\
	\eea
	or in terms of  $x=\Omega_{N-1}^{-1/2}z$,
	\bea
	\rho(z_a,z_b)=e^{-\frac{1}{2}|z_a|^2-\frac{1}{2}|z_b|^2-\frac{1}{4}(z_a+z_b)^{T}\beta_1(z_a+z_b)}\nonumber\\
	\label{B11}
	\eea
	with 
	\be
	\beta_1=\bigg(N\,\tan \frac{\pi}{2N}\bigg)\bigg(\Omega_{N-1}^{-\frac{1}{2}}\beta \beta^{T}\Omega_{N-1}^{-\frac{1}{2}}\bigg)
	\ee
	To obtain the entanglement entropy  for this simple case, we need to diagonalize $\beta_1$. The result is one finite eigenvalue $\chi_1=N\,\tan \frac{\pi}{2N}\times \beta^{T}\Omega_{N-1}^{-1}\beta$
	and $N-2$ zero eigenvalues.  Since
	\bea
	\beta^{T}\Omega_{N-1}^{-1}\beta=-\frac{1}{N}\sum_{m=2}^{N}\beta_m=\frac{1}{N}\beta_1=\frac{1}{N}\cot \frac{\pi}{2N}\nonumber\\
	\eea
	it follows that $\chi_1=1$. Therefore, the density matrix (\ref{B11}) simplifies
	\be
	e^{-\frac{y_1^2}{2}-\frac{y_2^2}{2}-\frac{\chi_1^2(y_1+y_2)^2}{4}}\rightarrow e^{-\frac{1}{2}(y_1-y_2)^2}
	\ee
	as expected from translational invariance. This case should therefore subtracted,
	with the (subtracted) entanglement entropy ${\cal S}_E(n=1,N)=0$, whatever $N$.

	\subsection{General case}\noindent
	For general $n>1$, we define the square matrix $\tilde\beta$ associated  to (\ref{DIS8})
	\bea
	\label{DIS9}
	\Omega_{N,n}^{-\frac 12}\tilde\beta_{N,n}\Omega_{Nn}^{\frac 12}\equiv 
	\Omega_{N,n}^{-\frac 12}B_{N,n}\Omega^{-1}_{N,N-n}B^T_{Nn}
	\Omega_{N,n}^{-\frac 12}
	\eea
	with the formal eigenvalue spectrum
	\bea
	\label{DIS10}
	\tilde\beta_{N,n}V_{N,n}^i=\chi_{N,n,i}V^i_{N,n}
	\eea
	and $i=1,..,n$. In terms of (\ref{DIS10}) the eigenvalues of the entangled density matrix (\ref{DIS9}) are
	\bea
	\label{DIS11}
	p_l[N,n,i]=(1-\lambda_{N,n,i})\lambda^l_{N,n,i}
	\eea
	with
	\bea
	\label{DIS12}
	\lambda_{N,n,i}=\frac{\chi_{N,n,i}/2}
	{1-\chi_{N,n,i}/2+\sqrt{1-\chi_{N,n,i}}}
	\eea

	\begin{figure*}
		\centering
		\includegraphics[width=0.48\linewidth]{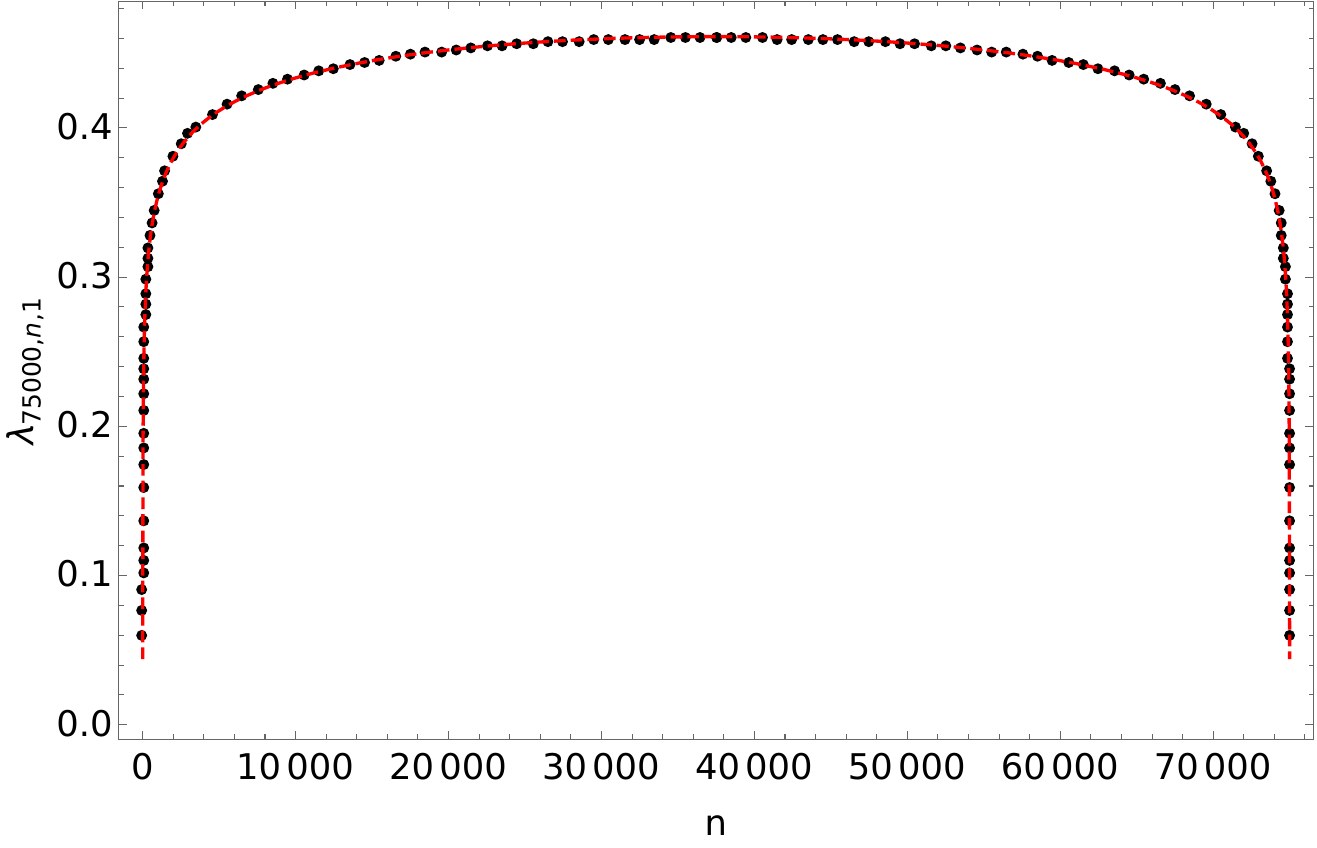}
		\includegraphics[width=0.48\linewidth]{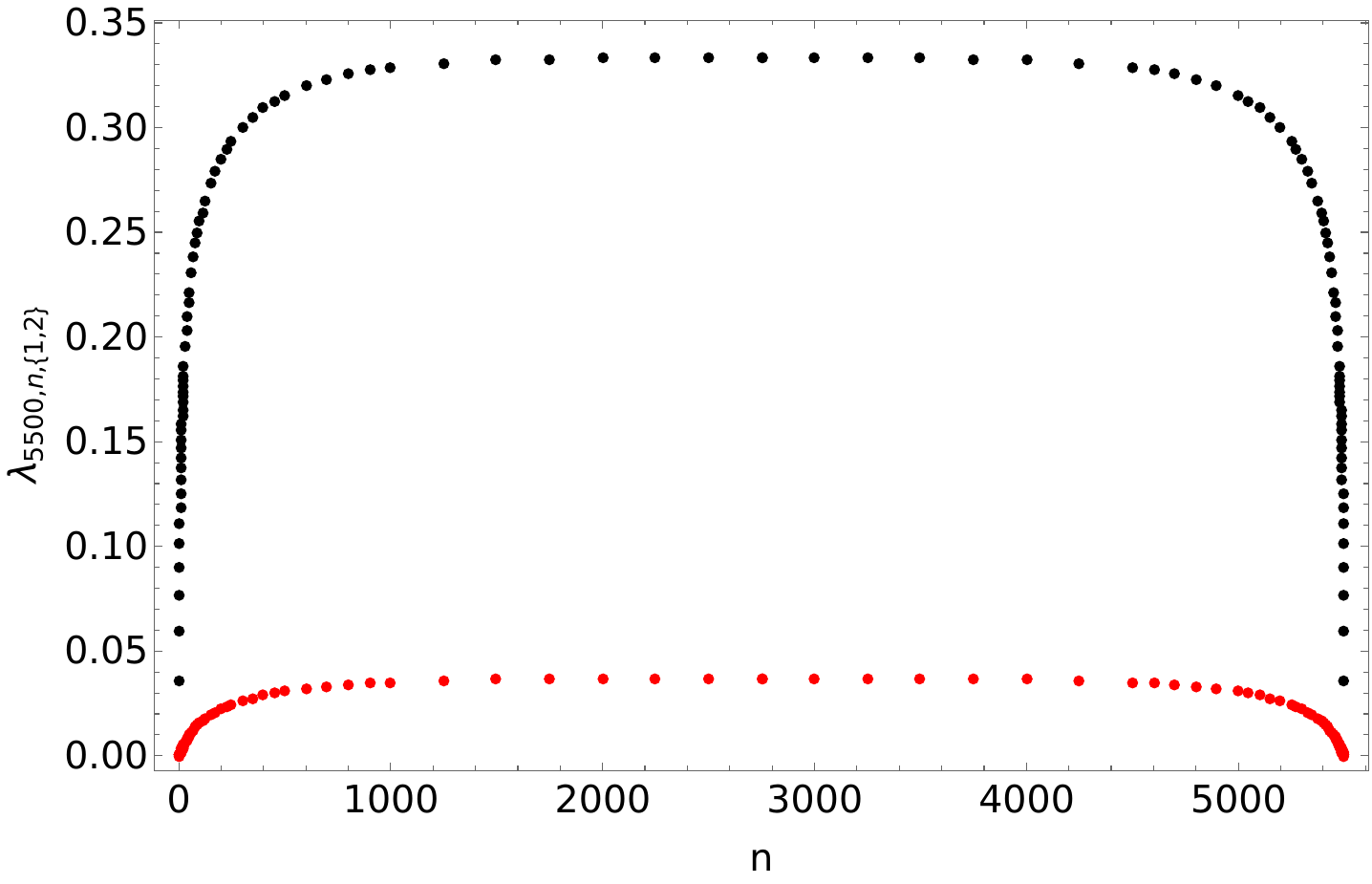}
		\includegraphics[width=0.48\linewidth]{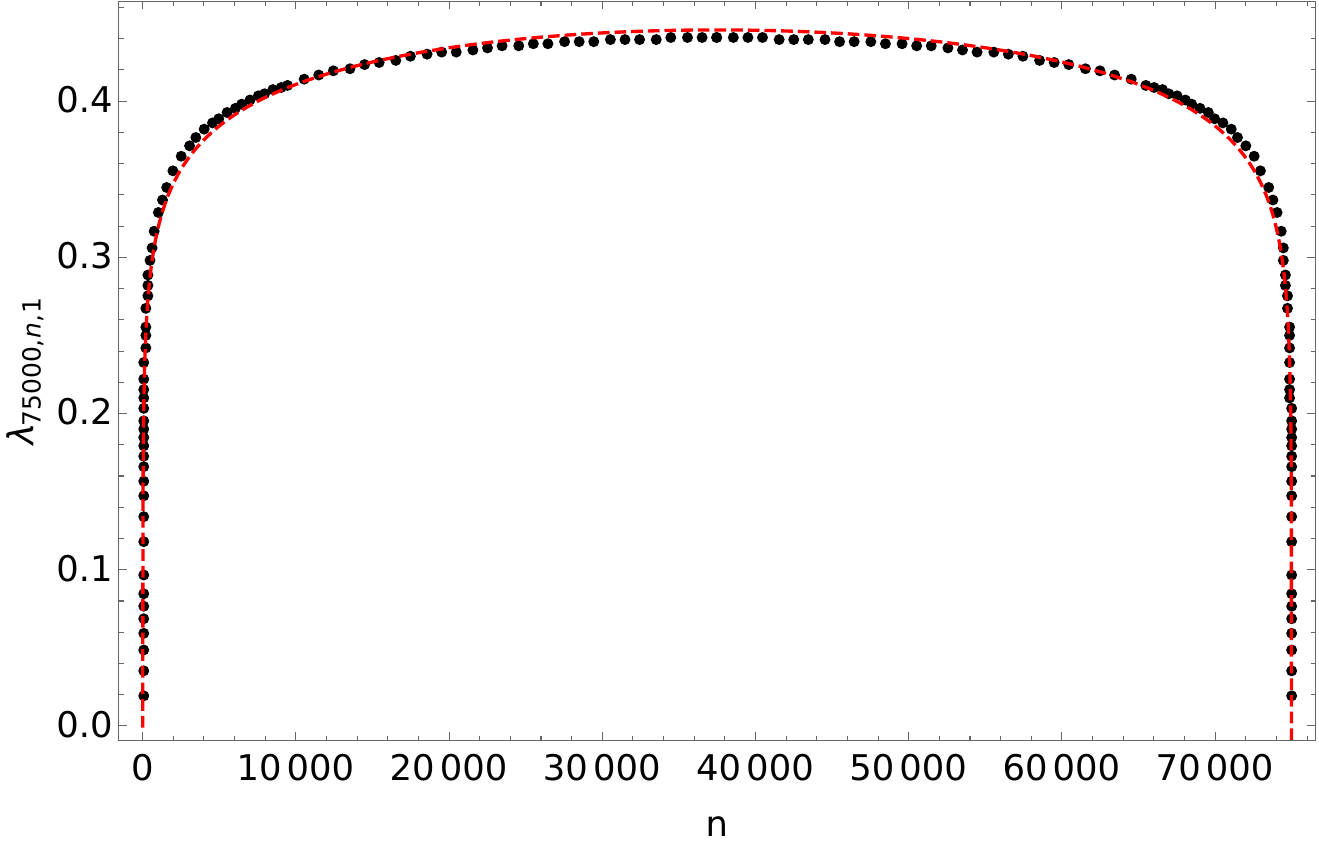}
		\includegraphics[width=0.48\linewidth]{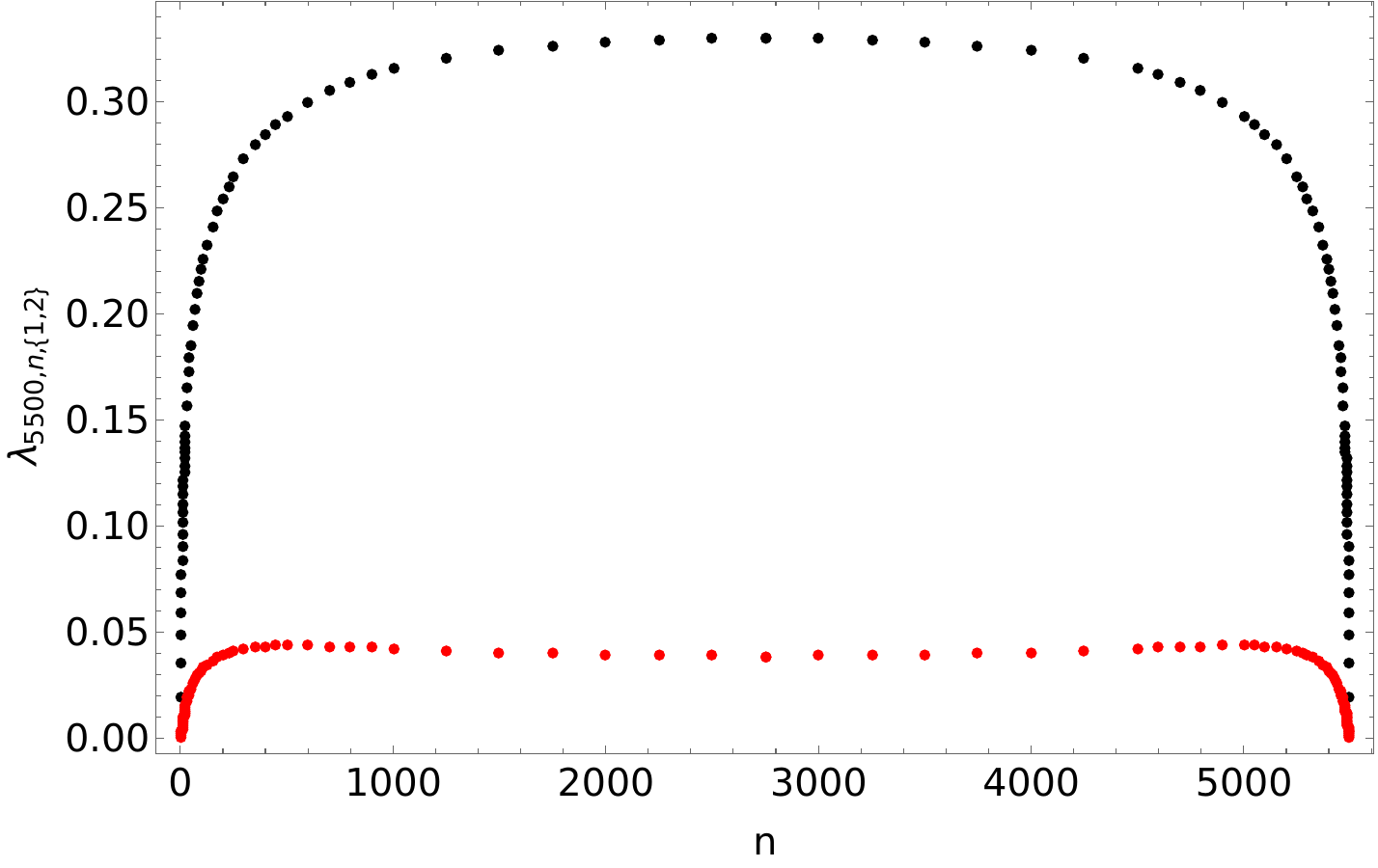}
		\caption{Lowest eigenvalue versus the blocked length $n$,  for a massless  open chain (top-left),  a massive open chain (top-right), 
			a massless closed chain (bottom left) and a massive closed chain (bottom right). See text.
		}
		\label{fig:LOWESTEV}
	\end{figure*}

	\subsection{Spectral flow of the largest eigenvalue}\noindent
	Most of the eigenvalues of the entangled density matrix  except one $\lambda_{N,n,1}$, are exponentially small and randomly distributed following a Poisson distribution. This observation is consistent with the results in~\cite{Liu:2018gae} for the massless case. 
	The exception  is a large and collective eigenvalue, 
	\bea
	\label{FLOW1}
	\lambda_{N,n,1}\approx 
	\bigg(1-e^{-S(am_\theta(T),n)}\bigg)
	\eea
	or equivalently
	\bea
	\label{FLOW2}
	p_l[N,n,1]\approx e^{-S(am_\theta(T), n) }
	\bigg(1-e^{-S(am_\theta(T), n) }\bigg)^l\nonumber\\
	\eea
	To derive \eqref{FLOW2}, we retained the largest eigenvalue  in \eqref{SEEX} only, and assumed that  $S_E$ is large, i.e.
	\begin{align}
	&\bigg({\rm ln}(1-\lambda_{N,n,1})+
	\frac{\lambda_{N,n,1}}{1-\lambda_{N,n,1}}\,{\rm ln}\,\lambda_{N,n,1}\bigg)\nonumber\\&=S_E+(e^{S_E}-1)\log(1-e^{-S_E})\approx S_E.\nonumber
	\end{align}
	The largest eigenvalue (\ref{FLOW2})  obeys the 
	cascade equation
	\bea
	\frac{dp_l}{d{\rm ln}n}=\frac 12
	C(m_\theta(T) na)(-p_l+lp_{l-1})
	\eea
	with a rate fixed by the entropic function of $\frac 12$-Dirac massive fermion in (\ref{CML1}).

	The exact form of the bosonic entropy $S(am_\theta(T))$ in the
	continuum is not known analytically, although its entropic derivative (\ref{B2}) for large or small cuts is. It is also UV sensitive. As we suggested earlier, the central charge for free massive bosons,  is to a good approximation analogous to the central charge of  $\frac 12$-Dirac (Majorana) massive fermions  as derived in~\cite{Liu:2022qqf}. Modulo a shift $\Lambda'$, we  have
	\begin{widetext}
		\bea
		\label{MAJORANA}
		S(am_\theta(T), n)\approx \frac 16\int_0^1dx
		\bigg(K_0\bigg(\frac{am_\theta(T)}{\sqrt{x(1-x)}}\bigg)-
		K_0\bigg(\frac{nam_\theta(T)}{\sqrt{x(1-x)}}\bigg)\bigg)\ 
		\eea
	\end{widetext}

	In Fig.~\ref{fig:LOWESTEV}
	top-left,  we show the results for the lowest eigenvalue of the open and massless chain, with $N=75000$.  The red dashed line is a fit to 
	\begin{equation}
	\lambda_{N,n,1}=  c_1\!\left(\!1-\exp\!\left(\!-c_2 \log \!\left(\frac{N \sin \left(\frac{\pi  n}{N}\right)}{\pi }\right)\!\right)\right.\label{eq:fit}
	\end{equation}
	with $c_1=0.829$ and $c_2=0.081$. Even though the result converges toward $c_2=1/6$, $N$ is still not large enough to be in the asymptotic regime, see Fig.~\ref{fig:SEE-MASS0}. In Fig.~\ref{fig:LOWESTEV}
	top-right, we show the results for the lowest two eigenvalues of an open and massive chain, with $N=5500$ and $m_\theta=0.001$. 
	
	In Fig.~\ref{fig:LOWESTEV} bottom-left, we show the results for the massless and  closed chain, with $N=75000$. 
	The red dashed line is a fit to \eqref{eq:fit} with $N=50000$, 
	$c_1=1.819$ and $c_2=0.02786$. In Fig.~\ref{fig:LOWESTEV}
	bottom-right, we show the two lowest eigenvalues for a closed and massive chain with  $N=5500$ and $m_\theta=0.001$. The spurious translational eigenvalue detailed in Appendix~\ref{DETAILS},  has been removed.

	\bibliography{references}
\end{document}